\documentclass[twocolumn]{aastex63}
\usepackage[T1]{fontenc}
\usepackage[utf8]{inputenc}
\usepackage{amsmath}
\usepackage{tabularx, booktabs}
\usepackage{graphicx}
\usepackage{natbib}
\usepackage[scaled]{helvet}
\usepackage{epsfig}
\usepackage{url}
\bibpunct{(}{)}{;}{a}{}{,}
\usepackage{textcomp}
\usepackage{mathpazo}
\usepackage{xspace}
\usepackage{apjfonts}
\usepackage{rotating}
\usepackage{gensymb}
\usepackage{units}

\usepackage{color}
\usepackage[normalem]{ulem}

\newcommand\vsini{\ifmmode{v\sin{i_\star}}\else $v\sin{i_\star}$\fi}
\newcommand\sini{\ifmmode{\sin{i_\star}}\else $\sin{i_\star}$\fi}

\newcommand{\msun}{\ensuremath{\,M_\Sun}}
\newcommand{\rsun}{\ensuremath{\,R_\Sun}}

\newcommand{\mjup}{\ensuremath{\,M_{\rm J}}}
\newcommand{\rj}{\ensuremath{\,R_{\rm J}}}
\newcommand{\rjup}{\ensuremath{\,R_{\rm J}}}
\newcommand{\rearth}{\ensuremath{\,R_{\rm \Earth}}\xspace}
\newcommand{\mearth}{\ensuremath{\,M_{\rm \Earth}}\xspace}

\newcommand{\kep}{{\it Kepler}}

\newcommand{\tess}{{\it TESS}}

\newcommand\mysim{\mathord{\sim}}
\newcommand{\kms}{\,km\,s$^{-1}$}
\newcommand{\ms}{\,m\,s$^{-1}$}

\newcommand{\rffigl}[1]{Figure~\ref{fig:#1}}

\newcommand{\rfsec}[1]{\mbox{\S\ \ref{sec:#1}}}
\newcommand{\rfsecl}[1]{\mbox{Section \ref{sec:#1}}}
\newcommand{\rfsecs}[2]{\mbox{Sections \ref{sec:#1} and \ref{sec:#2}}}

\newcommand{\rftabl}[1]{Table~\ref{tab:#1}}

\newcommand{\cfa}{Center for Astrophysics \textbar \ Harvard \& Smithsonian, 60 Garden St, Cambridge, MA 02138, USA}
\newcommand{\boulder}{Department of Astrophysical and Planetary Sciences, University of Colorado, Boulder, CO 80309, USA}

\newcommand{\wisconsin}{Department of Astronomy, University of Wisconsin─-Madison, 475 North Charter Street, Madison, WI 53706, USA}
\newcommand{\amnh}{Department of Astrophysics, American Museum of Natural History, 200 Central Park West, Manhattan, NY, USA}
\newcommand{\cca}{Center for Computational Astrophysics, Flatiron Institute, 162 5th Avenue, Manhattan, NY, USA}

\newcommand{\hubble}{\altaffiliation{NASA Hubble Fellow}}

\DeclareUnicodeCharacter{2500}{\mbox{\vrule height2.2ptdepth-1.8ptwidth.5em}}

\begin{document}

\title{A decade of radial-velocity monitoring of Vega and new limits on the presence of planets}

%Authors
\correspondingauthor{Spencer Hurt}%Agreed
\email{spencer.hurt@colorado.edu}

\author[0000-0002-6903-9080]{Spencer A. Hurt}
\affiliation{\boulder}

\author[0000-0002-8964-8377]{Samuel N. Quinn}
\affiliation{\cfa}

%%%%%%%%%%%%%%%%%%%%%%%%%%%%%%%%%%%%%%%%%%%%%%

\author[0000-0001-9911-7388]{David W. Latham} %responded, made changes
\affiliation{\cfa}

\author[0000-0001-7246-5438]{Andrew Vanderburg} %responded, made changes
\affiliation{\wisconsin}

\author[0000-0002-9789-5474]{Gilbert A. Esquerdo} 
\affiliation{\cfa}

\author[0000-0002-2830-5661]{Michael L. Calkins} %responded, no changes suggested
\affiliation{\cfa}

\author{Perry Berlind}
%responded, made changes
\affiliation{\cfa}

\author[0000-0003-4540-5661]{Ruth Angus}
%responded, made changes
\affiliation{\amnh}
\affiliation{\cca}

\author{Christian A. Latham}
\affiliation{\cfa}

\author[0000-0002-4891-3517]{George Zhou} 
%responded, no changes
\hubble
\affiliation{\cfa}

\shorttitle{Vegan planets and stellar activity}
\shortauthors{Hurt et al.}

\begin{abstract}
We present an analysis of $1524$\ spectra of Vega spanning $10$\ years, in which we search for periodic radial velocity variations. A signal with a periodicity of $0.676$\ days and a semi-amplitude of $\mysim10$\ \ms\ is consistent with the rotation period measured over much shorter time spans by previous spectroscopic and spectropolarimetric studies, confirming the presence of surface features on this A0 star. The timescale of evolution of these features can provide insight into the mechanism that sustains the weak magnetic fields in normal A type stars. Modeling the radial velocities with a Gaussian process using a quasi-periodic kernel suggests that the characteristic spot evolution timescale is $\mysim180$\ days, though we cannot exclude the possibility that it is much longer. Such long timescales may indicate the presence of failed fossil magnetic fields on Vega. \tess\ data reveal Vega's photometric rotational modulation for the first time, with a total amplitude of only $10$\ ppm, and a comparison of the spectroscopic and photometric amplitudes suggest the surface features may be dominated by bright plages rather than dark spots. For the shortest orbital periods, transit and radial velocity injection recovery tests exclude the presence of transiting planets larger than $2$\ \rearth\ and most non-transiting giant planets. At long periods, we combine our radial velocities with direct imaging from the literature to produce detection limits for Vegan planets and brown dwarfs out to distances of $15$\ au. Finally, we detect a candidate radial velocity signal with a period of $2.43$\ days and a semi-amplitude of $6$\ \ms. If caused by an orbiting companion, its minimum mass would be $\mysim20$\ \mearth; because of Vega's pole-on orientation, this would correspond to a Jovian planet if the orbit is aligned with the stellar spin. We discuss the prospects for confirmation of this candidate planet.

\vspace{.85cm}
\end{abstract}

\section{Introduction}

The search for exoplanets has traditionally focused on low-mass (FGKM) stars, as intermediate mass stars pose several observational challenges. For example, their larger size and mass translate to smaller transit and radial velocity signals for a given planet size, and rapid rotation and reduced radial velocity information content prevent precise Doppler spectroscopy \citep[e.g.,][]{Beatty:2015}. However, careful target selection and expanded observational techniques have helped to overcome these difficulties and have led to new opportunities for planet characterization. When A-type stars leave the main sequence, they cool and spin down, providing RV surveys a means to explore planet populations around intermediate mass stars. Observations of post-main sequence stars show that `retired A stars' are more likely than Sun-like stars to host massive planets \citep{Johnson20102,Johnson20101,Ghezzi2018}. However, these analyses suggest a low occurrence rate of planets on close-in orbits. Data from NASA's Transiting Exoplanet Survey Satellite \citep[\tess;][]{Ricker:2015}, reveal that the same is true of A-type main sequence stars: the occurrence rate of hot Jupiters orbiting A stars is low, and not dissimilar to that of hot Jupiters orbiting solar-type stars \citep{zhou:2019}.

A-type stars are often over-represented in wide-field transit surveys due to their intrinsic brightness, so despite the challenges of detecting and characterizing hot Jupiters orbiting A stars, many have now been discovered.  While A-type stars are often rapidly rotating and this poses problems for mass measurements, it can facilitate characterization sometimes not possible for other stars. Stellar obliquity, for example, can be measured from spectroscopic transit observations or transits across a gravity darkened stellar surface. For hot Jupiter hosts above the Kraft break \citep{Kraft:1967}, spin-orbit misalignment is often observed \citep[such that stellar obliquities appear to be consistent with an isotropic distribution;][]{Winn:2010b,Schlaufman2010,Albrecht:2012}, and A stars appear to be no exception \citep[e.g.,][]{CollierCameron:2010,zhou:2016,zhou:2019,Ahlers2020a,Ahlers2020b}. It is unclear if this is due to primordial misalignment or orbital migration and whether this stellar obliquity extends to small planets or long period planets.

Due to their intrinsic brightness, A stars are good targets for imaging surveys, for which the brightness of the stars is among the primary concerns. A stars are also intrinsically young (since their main-sequence lifetimes are short), which enhances the likelihood of detecting debris disks before they disperse or self-luminous planets and brown dwarfs before they cool. Vega is a $0$\textsuperscript{th} magnitude A0V star (see \rftabl{stellarparams} for additional stellar parameters) and the anchor of the Vega magnitude system. Given its brightness and its special history as a spectrophotometric calibrator \citep[see, e.g.,][]{hayes:1975}, it is a particularly well observed star. 

Ever since the IRAS discovery of a circumstellar disk \citep{Aumann1984}, Vega has been a frequent target of imaging studies. Though early detections of the dust around Vega appeared to show a clumpy, asymmetrical formation \citep{Holland1998, Koerner2001, Wilner2002, Marsh:2006}, more recent data reveal that the disk is smoother than originally thought \citep{Su2005, Sibthorpe2010, Pietu2011, Hughes2012, Holland2017}. ALMA observations have resolved the structure of Vega's outer dust belt, which extends to $150$--$200$\ au and has a steep inner edge at $60-80$\ au \citep{Matra:2020}. \textit{Spitzer} observations detect mid-IR excess in the disk consistent with an asteroid belt located at $14$\ au \citep{Su2013}. And near-IR excess has been detected close to the star, corresponding to hot dust \citep{Absil:2006, Defrere2011}.

\begin{table}[!t]
\centering
\caption{Stellar Parameters of Vega}
\setlength{\tabcolsep}{5.9pt}
\begin{tabularx}{8cm}{@{} lrrr @{}}
\toprule
Parameter & Value & Units & Source \\
\midrule
R.A. & $18:36:56.336$ & hh:mm:ss & (1) \\
Dec. & $38:47:01.280$ & dd:mm:ss & (1) \\
$\mu_\alpha$ & $200.94$ & mas yr$^{-1}$ & (1) \\
$\mu_\delta$ & $286.23$ & mas yr$^{-1}$ & (1) \\
Parallax & $130.23 \pm 0.36$ & mas & (1) \\
Distance & $7.68 \pm 0.02$ & pc & (1) \\
Inclination & $6.2 \pm 0.4$ & degrees & (2) \\
Rotational period & $0.71 \pm 0.02$ & days & (2) \\
$T\textsubscript{eff}$ (Apparent) & $9660 \pm 90$ & K & (2) \\
$T\textsubscript{eff}$ (Pole) & $10070 \pm 90$ & K & (2) \\
$T\textsubscript{eff}$ (Equator) & $8910 \pm 130$ & K & (2) \\
$R\textsubscript{pol}$ & $2.418 \pm 0.012$ & \rsun & (2) \\
$R\textsubscript{eq}$ & $2.726 \pm 0.006$ & \rsun & (2) \\
Mass & $2.15 \substack{+0.10 \\ -0.15}$ & \msun & (2) \\
Upper Age Estimate & $700 \substack{+150 \\ -75}$  & Myr & (2) \\
Lower Age Estimate & $471 \pm 57$  & Myr & (3) \\
\bottomrule
\end{tabularx}
\\[1ex]
Notes: (1) \cite{vanLeeuwen:2007}; (2) \cite{Monnier2012}; (3) \cite{Yoon:2010}
\label{tab:stellarparams}
\end{table}

While \cite{Yelverton2020} show that there are no clear or strong planet-debris disk correlations, systems such as HR $8799$ \citep{Marois2008, Marois2010}, $\beta$ Pic \citep{Lagrange2009, Lagrange2010}, and 51 Eridani \citep{Macintosh2015} provide examples of stars hosting both imaged planets and a disk. Furthermore, features in a disk can be used to investigate the possible presence of planets and their properties, and Vega's disk is a complex system that contains features that could arise from a planetary system. For a star of Vega's age, circumstellar disks are maintained by debris from colliding planetesimals \citep{Wyatt2008}. The warm and cold belts are potential sources for this debris, but radiation pressure, stellar winds, and Poynting Robertson drag forces mean a high dust production rate is necessary to maintain the disk. \cite{Defrere2011} conclude that major dynamic perturbations are necessary to produce the quantities of dust observed, suggesting a system of giant planets migrating outwards. \cite{Raymond2014} use dynamical simulations to find that low-mass, closely-spaced planets could efficiently scatter exocomets inwards, accounting for the hot dust. A possible configuration includes a system of planets between $5$ and $60$\ au, wherein the outermost planets have masses less than about $20$\ \mearth, and they suggest that a Jupiter mass planet beyond $15$\ au would disrupt the inward-scattering chain.

Gaps in the disk can also be used to infer the presence of planets. Observed depletion of dust between detected belts has been suggested to indicate the presence of multiple long-period giant planets \citep{Su2013}, a single $3$ \mjup\ planet at $75$\ au \citep{Zheng2017}, or a chain of low-mass planets or a Saturn-mass planet \citep{Bonsor2018}. 
\cite{Matra:2020} argue that the steep inner edge of the cold belt cannot be explained by collisional evolution, which would result in a shallow inner slope. It could be explained by a chain of planets in which the outermost is located near $70$\ au and has a mass greater than $6$ \mearth, or by a single planet with a mass of $5$ \mjup\ and an orbital radius of $50-60$\ au. Clearly not all of these architectures can be present simultaneously, but the hot dust, gap structures, and characteristics of the cold belt all allow for the existence of a system of planets.

\begin{figure*}[!ht]
\centering\includegraphics[width=\linewidth]{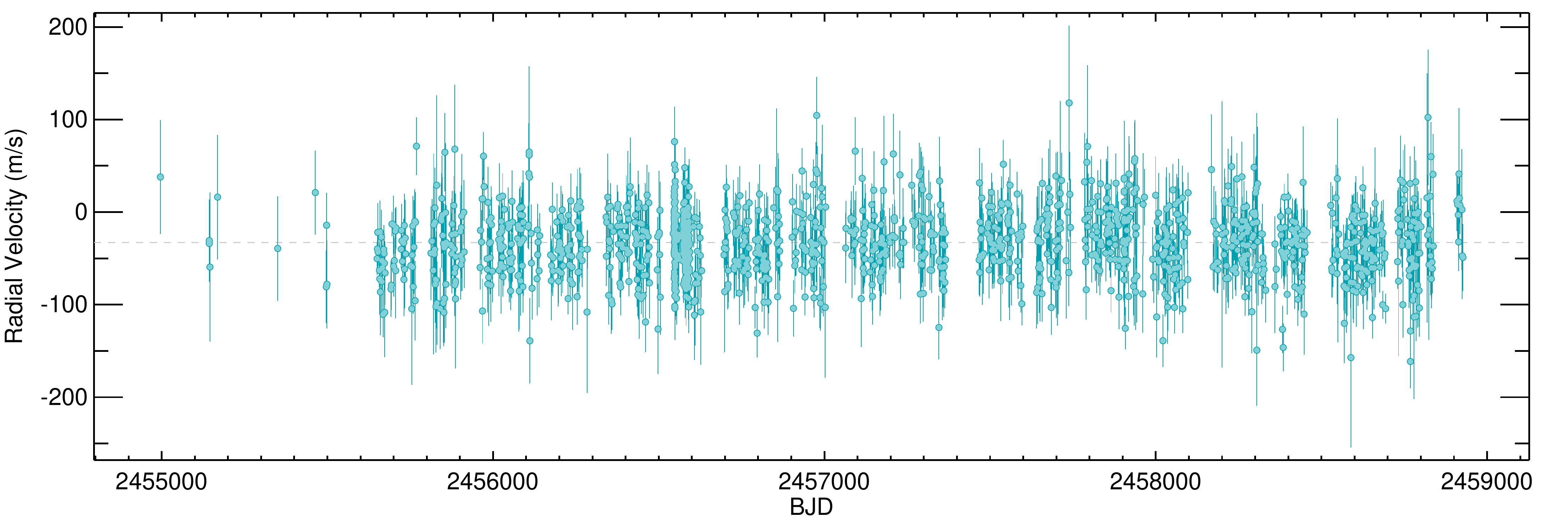}\caption{Relative radial velocities of Vega, derived from TRES spectra.}
\label{fig:fullrv}
\end{figure*}

We view Vega approximately pole-on, allowing a coplanar planet on a near-circular orbit to always be observed near its maximum separation, which further improves Vega's suitability for direct imaging. In the most recent published search, \cite{Meshkat2018} explored the inner $15$\ au using the P1640 instrument on Palomar's 5.1-m Hale telescope and placed mass detection limits of $\mysim20-85$\ \mjup. \cite{Heinze2008} provide MMT observations to constrain objects between $20$\ and $80$\ au, finding an upper limit of $\mysim5-20$\ \mjup. And with \textit{Spitzer} observations, \cite{Janson2015} rule out any planets with a mass greater than $\mysim1-4$\ \mjup\ between $100$ and $200$\ au. These represent the lowest mass limits in the literature for objects orbiting Vega. 

\begin{table}[!b]
\centering
\caption{TRES Radial Velocities of Vega}
\setlength{\tabcolsep}{0.95cm}
\begin{tabularx}{0.45\textwidth}{@{} lrr @{}}
\toprule
~BJD & RV~~ & $\sigma$~~~~\\
     & (\ms) & (\ms)    \\
\midrule
~2456025.026123      &   -10.0   &    23.1 \\
~2456026.017315      &   -20.8   &    22.8 \\
~2456027.016471      &   -16.1   &    25.4 \\
~2456028.020389      &   -48.7   &    24.9 \\
~2456029.020685      &    17.2   &    24.4 \\
~2456030.020025      &   -69.8   &    21.3 \\
~2456031.019574      &   -37.6   &    19.8 \\
~2456033.007404      &   -59.8   &    19.7 \\
~2456034.010060      &     2.9   &    16.0 \\[0.8ex]
\bottomrule
\end{tabularx}
\tablecomments{The full set of velocities is available as a machine-readable table. A portion is shown here for form and content.\\[-1ex]}
\label{tab:rvtable}
\end{table}

The inner working angles of direct imaging instruments prevent good limits on the existence of planets for scales similar to or smaller than the inner solar system. Radial velocities can be used to complement imaging at small separations and for somewhat lower masses at large separations. The problem of rapid rotation remains, of course. Spectropolarimetric observations measure a rotation period of $0.678\substack{+0.036 \\ -0.029}$\ days \citep{Alina2012}. A spectroscopic analysis by \cite{Bohm2015} indicates stellar activity modulated at the same period. These results suggest a stellar rotational velocity of nearly $200$\ \kms, but the pole-on orientation leads to projected rotation of only about $20$\ \kms, which does not severely limit radial velocity precision. Indeed, velocities from \cite{Bohm2015} display scatter on the order of only 10 \ms. One concern is that because of the stellar orientation, radial velocities will be most sensitive to planets misaligned with the star. However, at least for giant planets on short periods, misaligned orbits appear to be the rule rather than the exception for A-star hosts.

In this work, we present our analysis of $1524$ spectra of Vega, which can be used to study stellar activity and search for planets smaller than $1$\ \mjup\ on short periods and massive planets out to 15\ au. In Section \hyperlink{section.2}{2}, we present the observations and data reduction. In Section \hyperlink{section.3}{3}, we explore the radial velocities for periodic signals and discuss their origin. In Section \hyperlink{section.4}{4}, we calculate detection limits. Lastly, in Section \hyperlink{section.5}{5}, we discuss our results.

\section{Observations}

\subsection{TRES Spectroscopy}

We obtained high-resolution spectra of Vega with the Tillinghast Reflector Echelle Spectrograph \citep[TRES; ][]{furesz:2008}, which is mounted on the 1.5-m Tillinghast Reflector at Fred L. Whipple Observatory on Mount Hopkins, AZ. It has a resolving power of $R\mysim44,000$, and a wavelength coverage of $3850$--$9100$\ {\AA}. We obtained a total of 1524 spectra spanning the 10-year period between UT 2009 June 13 and 2019 October 24. Typical exposure times ranged from a few seconds to a few tens of seconds, achieving signal-to-noise ratios (SNR) between about 300 and 1000 per resolution element. 

\begin{figure*}[!t]
\centering
\includegraphics[width=\linewidth]{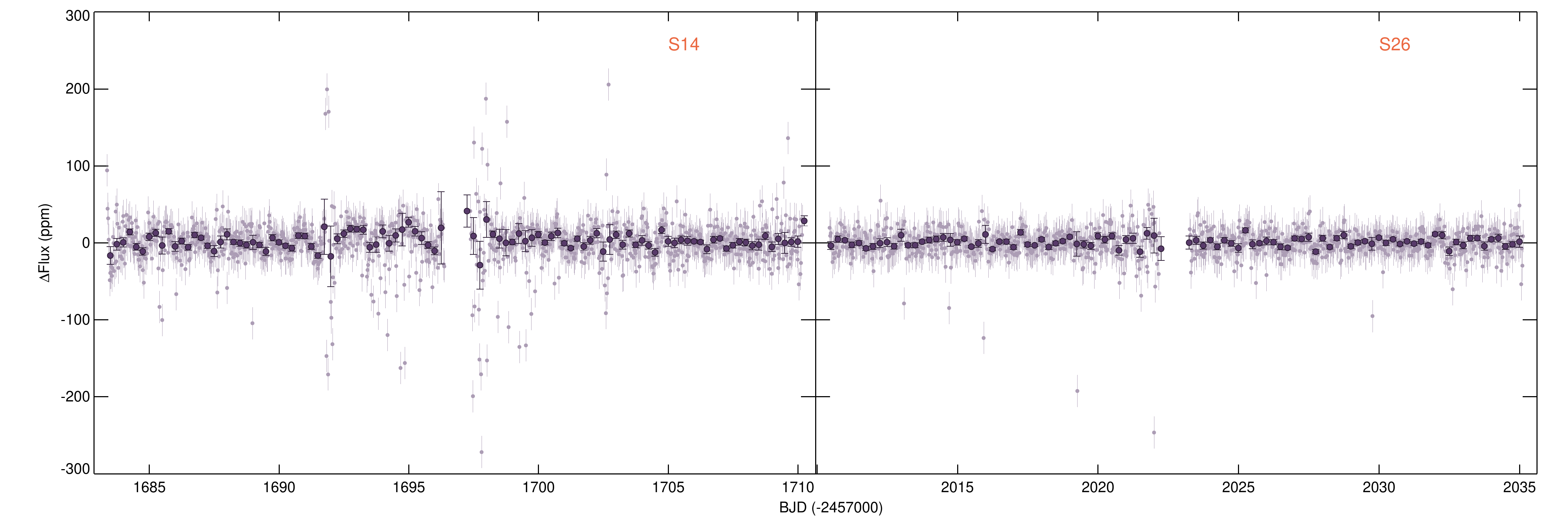}\\
\includegraphics[width=0.48\linewidth]{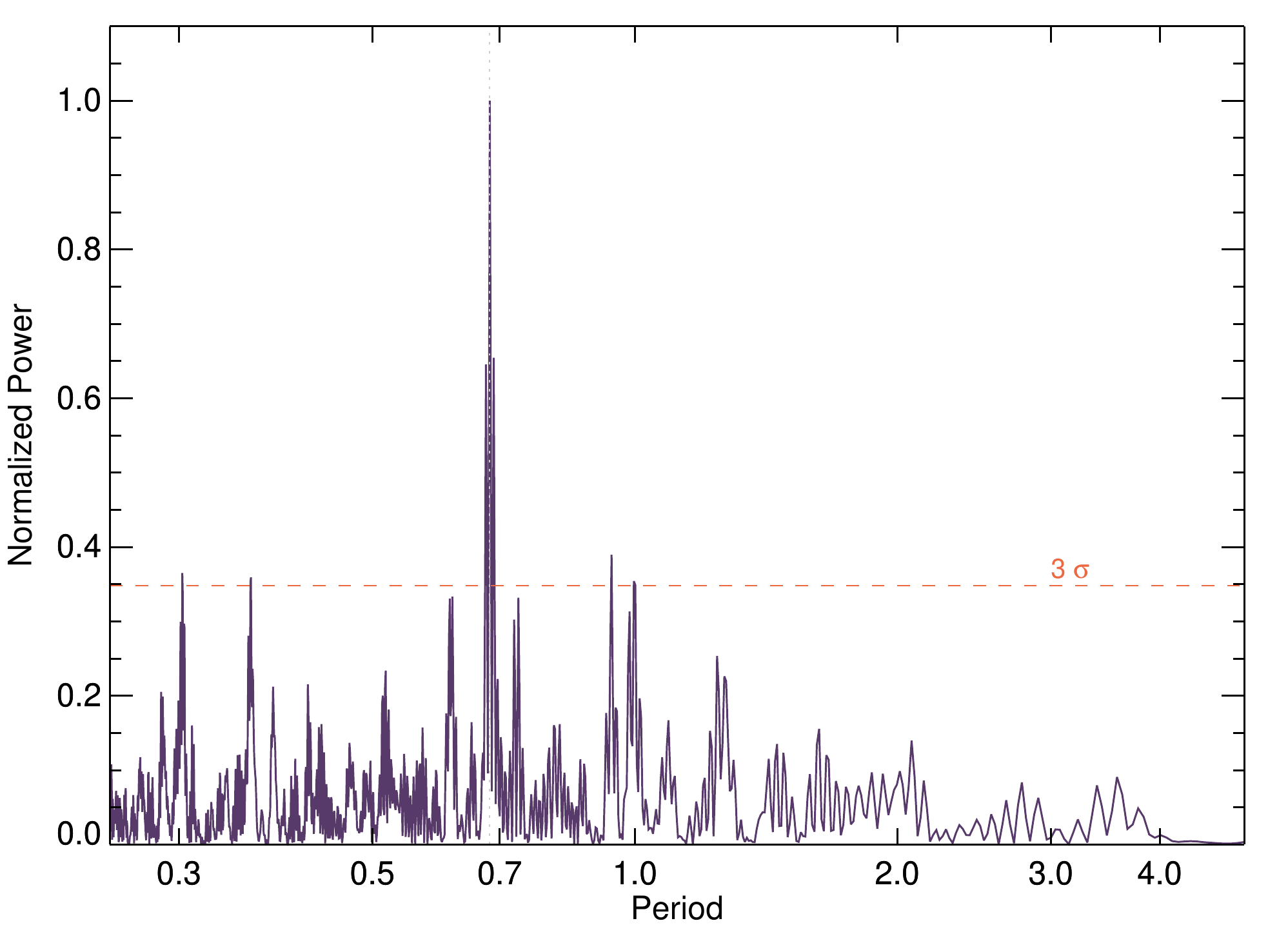}
\includegraphics[width=0.48\linewidth]{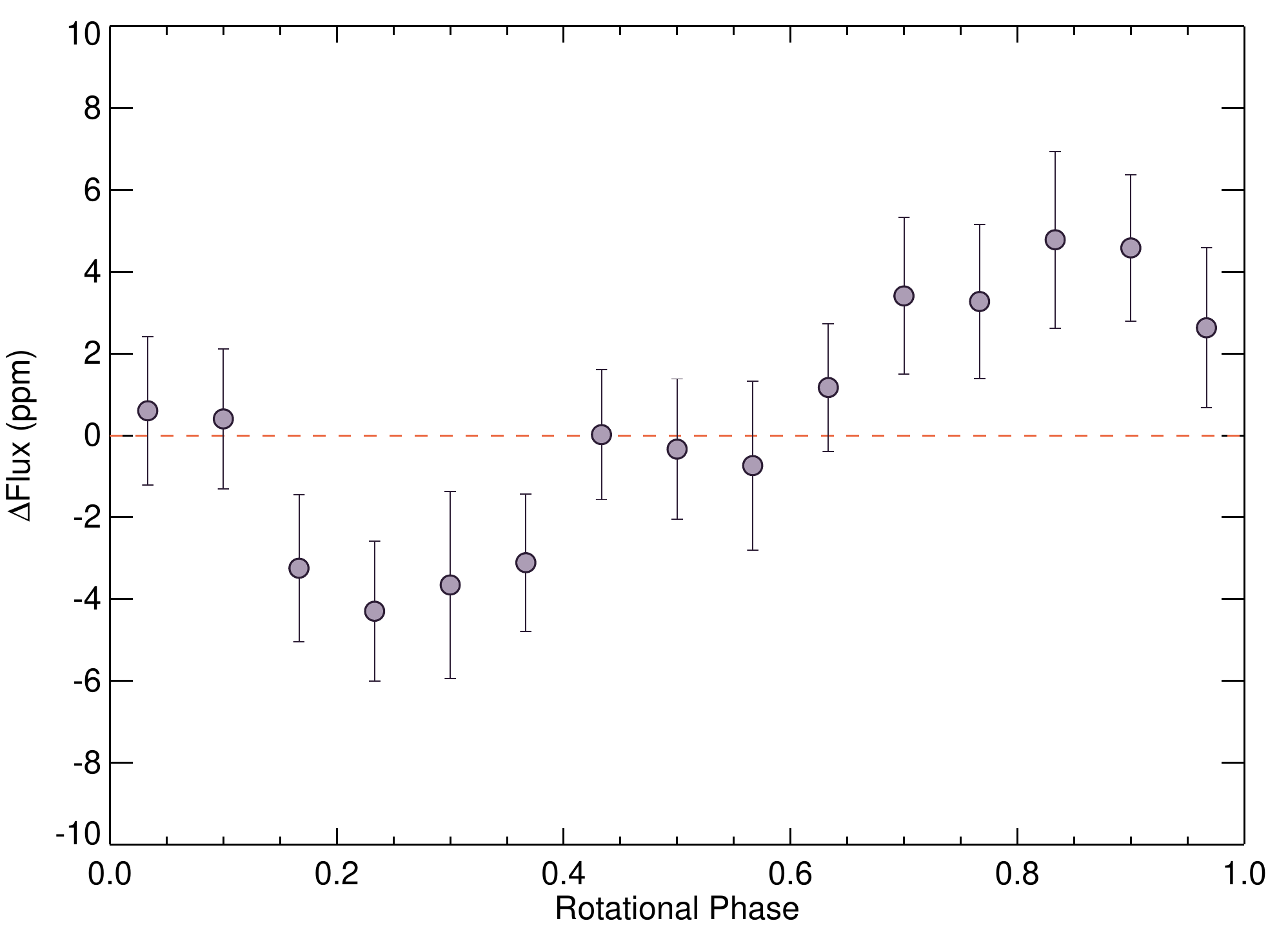}\\
\caption{Top: The \tess\ light curve of Vega, after decorrelation against the spacecraft quaternion time series, as described in \rfsec{tess}. Individual 30-minute cadence measurements are shown as small, light purple circles, while data in 6.5-hour bins are shown as large, dark purple circles. The median 6.5-hour standard deviation is 5.0 ppm. Lower Left: Lomb-Scargle periodogram of the \tess\ light curve, showing a peak at 0.68 days, consistent with the rotation period of the star. Lower Right: The binned, phase-folded \tess\ light curve, which shows a total variation of $\mysim 10$\,ppm.}
\label{fig:tesslc}
\end{figure*}

We optimally extracted and reduced the spectra following the procedures outlined in \citet{Buchhave:2010}, and while we begin by following the radial velocity measurement outlined therein, our final velocity extraction includes a few key differences. We first cross-correlate each spectrum of Vega against the strongest exposure, treating each spectral order separately. The relative RV for each exposure is taken to be the location of the peak of the summed CCF (across all orders) from that spectrum. The internal RV uncertainty for each observation is taken to be the standard deviation of the locations of the CCF peak of each order for that spectrum. Next, we shift and median combine the 1524 spectra to generate a master template spectrum. In the case of Vega, the template SNR is 26,000 per resolution element. Though exposures of Vega are typically short enough that cosmic ray rejection is not very important, we identify outlier pixels and replace them with the median spectrum at that location. Finally, we re-run the order-by-order cross-correlation, this time against the high-SNR template. 

TRES was not designed for long-term stability at the level of meters per second, and has at times experienced drifts and jumps in its instrumental zero point as large as a few tens of \ms. Changes of this magnitude can mimic or mask the presence of long-period companions. To combat this problem, we track the zero point with nightly observations of several RV standard stars, allowing us to measure and correct for zero point changes over time. From the standard deviation of the RV standards, we also estimate the instrumental RV noise floor, to be added in quadrature with the internal error estimates described above. The TRES instrumental precision at the beginning of our data set was $\mysim 50$\ \ms, but by the end of 2010 had improved to $\mysim 10$--$15$\ \ms, thanks in large part to hardware upgrades. Though these introduced some zero point changes, they are corrected for in the same way as other zero point shifts.
Our final, zero-point-corrected, relative RVs are shown in \rffigl{fullrv} and presented in \rftabl{rvtable}.

\subsection{TESS Photometry}
\label{sec:tess}

Vega was observed by NASA's \tess\ mission \citep{Ricker:2015} in Sector 14, between UT 2019 July 18 and 2019 August 15, and in Sector 26, between UT 2020 June 8 and 2020 July 04. The star is bright enough to fill the full well depth of hundreds of pixels, but the \tess\ detector is designed to preserve the flux by spilling into neighboring pixels. As long as the star is not too close to the edge of the detector, the full frame images (FFIs), returned at 30-minute cadence, can be used to extract photometry from an area encompassing all of the light. The FFIs were processed by the Science Processing Operations Center (SPOC) at NASA Ames \citep{Jenkins:2015,Jenkins:2016}, adapted from the pipeline originally developed for the \kep\ mission \citep{jenkins:2010}. We then performed photometry using apertures shaped to trace the distribution of charge on the \tess\ images, including 3,625 pixels in Sector 14 and 4,335 pixels in Sector 26. We corrected the flux for sources within the aperture using \tess\ magnitudes listed in the \tess\ Input Catalog. To account for systematics caused by the motion of the spacecraft, we followed the procedure outlined in \citet{vanderburg:2019}. Namely, we assumed that systematics caused by changes in spacecraft pointing can be corrected by decorrelating against the background scattered light signal and combinations of the spacecraft quaternions, which encode the pointing of the spacecraft every 2 seconds. We produced quaternion time series by calculating the standard deviation and mean quaternions during each 30-minute exposure. Ultimately, we found the best performance (i.e., lowest resulting light curve scatter) when decorrelating against the scattered light signal and the first, second, and third order of these quaternion time series. We exclude a few hours of data most strongly affected by scattered light at the beginning of each orbit in Sector 26, and we use a spline to flatten remaining low frequency systematics that occur on the timescale of the spacecraft orbit.

\begin{figure*}[!htb]
\centering\includegraphics[width=\linewidth,trim={1.5cm 0 1.5cm 0},clip]{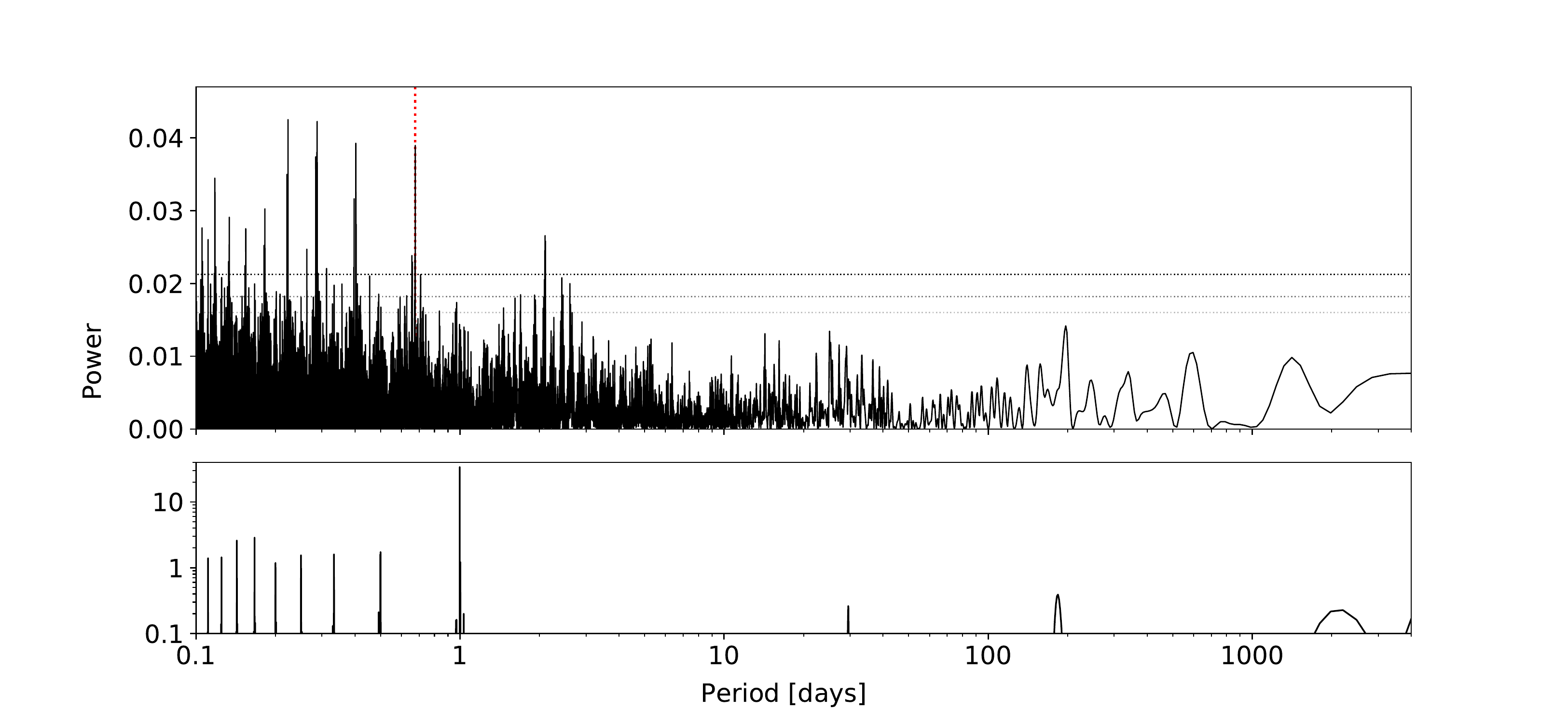}
\caption{GLS periodogram (top) and window function (bottom) for TRES radial velocities. The horizontal lines in the periodogram correspond to FAPs of $0.05$, $0.01$, and $0.001$. The peak located at $0.677$ days, corresponding the the rotational period, is marked in red.}
\label{fig:periodogram}
\end{figure*}

The resulting light curve is shown in \rffigl{tesslc}, and is remarkably quiet. The median standard deviation in 6.5-hour bins is only 5.0 ppm. We run {\tt Transit Least Squares} \citep[{\tt TLS},][]{Hippke:2019} in search of transiting planets but we find no evidence for transit-like features, and in \rfsecl{injection} produce detection limits using transit injection recovery. The maximum power in a Lomb-Scargle periodogram occurs for a period of 0.68 days (\rffigl{tesslc}, lower left panel), which is consistent with previous measurements of the rotation period of Vega. We do detect the signal in both sectors individually at lower significance, but because of the year-long gap between sectors we cannot confirm whether the signal is stable in phase for the full time span. It is interesting to note that while the period is still significant ($> 3 \sigma$) if we exclude photometric outliers, the significance is highest when they are included. This suggests that the apparent outliers vary in phase with the rest of the data and may be associated with stellar activity rather than systematics related to the instrument or data processing. This could indicate that some surface features are evolving in brightness on very short timescales, while others appear more stable over the course of a month, and perhaps over the full year spanned by the \tess\ data.

\section{Analysis of Radial Velocity Signals}
\label{sec:rvsignals}

In this section, we present analyses of the TRES data to search for and characterize radial velocity signals arising from rotation and orbiting companions. We begin our search for periodic signals in the radial velocity data using a generalized Lomb Scargle periodogram (GLS, \citealt{Zechmeister:2009}) and the window function, shown in \rffigl{periodogram}. The periodogram contains many significant peaks with false alarm probabilities (FAPs) below $0.001$, most of which fall at short periods. The window function reflects our nightly observing cadence, with strong power at a sampling rate of $1\ \rm{day}^{-1}$. Given that the majority of signals in the periodogram fall below $1$ day, the true frequency of variability is likely above the Nyquist frequency of .5 days$^{-1}$, and the rest of the peaks are aliases. We discuss identification of the true peak below and then model this signal and others with Gaussian processes and Keplerians.

\begin{table}[!t]
\small
\centering
\caption{Significant Periodogram Signals (FAP < 0.1\%)}
\setlength{\tabcolsep}{9.8pt}
\begin{tabularx}{\linewidth}{@{} lcccc @{}}
\toprule
Period & Frequency & Notes & Amplitude & FAP \\
(days) & (days$^{-1}$) & & (\ms) & \\
\midrule
0.105 & 9.48 & $\mathrm{F_{rot}} + 8$ & 7.16 & 7.71e-6 \\
0.111 & 9.00 & \nodata & 7.12 & 2.49e-5\\
0.118 & 8.48 & $\mathrm{F_{rot}} + 7$ & 8.21 & 3.80e-8\\
0.134 & 7.48 & $\mathrm{F_{rot}} + 6$ & 7.60 & 3.37e-6\\
0.154 & 6.48 & $\mathrm{F_{rot}} + 5$ & 7.59 & 7.63e-6\\
0.182 & 5.48 & $\mathrm{F_{rot}} + 4$ & 7.58 & 1.06e-6\\
0.221 & 4.52 & $\mathrm{\left|F_{rot} - 6\right|}$ & 8.31 & 2.35e-8\\
0.223 & 4.48 & $\mathrm{F_{rot}} + 3$ & 9.15 & 8.23e-11 \\
0.263 & 3.80 & \nodata & 6.88 & 9.05e-5 \\
0.284 & 3.52 & $\mathrm{\left|F_{rot} - 5\right|}$ & 8.17 & 5.79e-9 \\
0.288 & 3.48 & $\mathrm{F_{rot}} + 2$ & 9.43 & 1.06e-10 \\
0.397 & 2.51 & $\mathrm{\left|F_{rot} - 4\right|}$ & 7.40 & 4.36e-7 \\
0.403 & 2.48 & $\mathrm{F_{rot}} + 1$ & 8.72 & 8.48e-10\\
0.677 & 1.48 & $\mathrm{F_{rot}}$ & 9.06 & 9.86e-10\\
2.10 & 0.478 & $\mathrm{F_{rot}} - 1$ & 7.29 & 2.31e-5 \\
\bottomrule
\end{tabularx}
\label{tab:signals}
\end{table}

\subsection{A Signal Arising from the Rotation of Active Regions}

\subsubsection{Detection of the Activity Signal}

One of the strongest peaks in the RV periodogram is located at $0.677$ days, which corresponds to the previously reported rotation period \citep{Alina2012,Bohm2015}. \rftabl{signals} lists all of the significant peaks in the periodogram, and shows how they relate to the rotation period. Nearly every signal is an alias of the $0.677$-day periodicity, with harmonics also present but at lower significance. While we have good reason to suspect that this value corresponds to the true signal, we conduct a more rigorous test of the aliases using the methodology of \cite{Dawson:2010}. This procedure simulates a sinusoidal signal at the candidate period using the time stamps of the observed data. The amplitudes and phases of each peak in the periodogram are then compared between the synthetic and actual data. The simulated signal that best reproduces the results from the observations is chosen as the true period. While the signal we observe is likely caused by activity, and therefore is somewhat irregular, this test does confirm $0.677$\ days as the best match to the observed periodogram.

\begin{figure}[!b]
\centering\includegraphics[width=\linewidth]{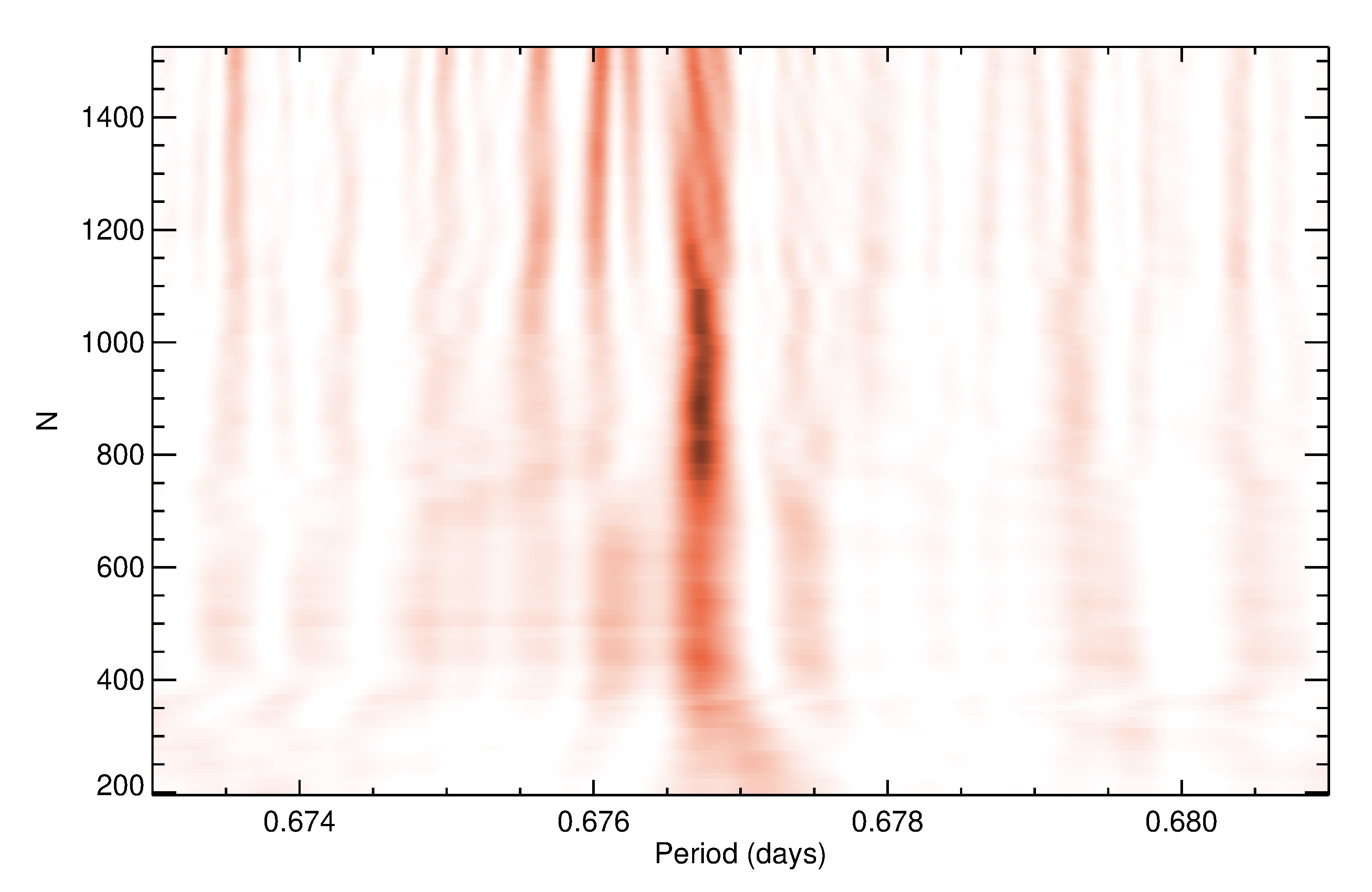}
\caption{Stacked periodogram of the TRES RVs, showing the periodicity at 0.677 days. The strength of the peak falls off around 1000 observations, after which multiple peaks emerge, suggesting that this signal originates from active surface regions rather than an orbiting companion.}
\label{fig:stacked}
\end{figure}

A useful diagnostic for distinguishing between activity and orbiting companions is the stacked periodogram, which is described by \cite{Mortier:2017}. As the number of observations included in a periodogram increases, the power of a peak corresponding to a planetary signal should monotonically increase, within the limits of the noise, while the power of a peak corresponding to rotation of features on the stellar surface may not, due to changes in the phase of the signal driven by the evolution of active regions. In \rffigl{stacked}, we present a stacked periodogram for our TRES radial velocities. For the first $\mysim 1000$ observations, the power at $0.677$ days increases, but afterwards, the power falls off. This supports the conclusion that the periodicity originates from activity, and its variation over the course of hundreds of observations is consistent with a long timescale of evolution.

\subsubsection{Modeling the Activity Signal}
\label{sec:model}

We first attempt to model the stellar activity with a Keplerian orbit using the {\tt radvel} Python package \citep{fulton:2018a}. However, Markov chain Monte Carlo (MCMC) samples failed to converge. Additionally, a periodogram of the residuals to the best fit showed significant power remaining near $0.677$\ days.  This suggests that a Keplerian model does not adequately fit the signal, which is expected given the structure seen in the stacked periodogram. It would have been surprising for the stellar surface to be described well by a single Keplerian (corresponding to a single active region) over the span of a decade.

We next consider a model in which active regions are very long-lived, but multiple regions may exist with slightly different periods due to differential rotation. 
For this exercise, we follow an iterative whitening procedure in which we fit a Keplerian with initial conditions determined from the periodogram following \citet{delisle:2016}. After refining that fit, we repeat the process with the strongest peak in the periodogram of the residuals, until there are no signals remaining with ${\rm FAP} < 0.1\%$. This procedure returns only two signals, $0.676085$\ and $0.676684$\ days, both coincident with the rotation period. This makes sense given that the same hemisphere of the star is almost always visible. Very low latitude (equatorial) active regions, visible for only a portion of the rotational phase, may appear at harmonics of the rotation period. Otherwise, variability caused by static surface features on Vega should be well described by a sum of Keplerians at the rotation period, with small differences in period caused by differential rotation. In \rftabl{activity} we present the parameters of the Keplerians used to model these two dominant signals. Some power remains in the periodogram of residuals near the rotation period, its aliases, and its harmonics, which may indicate the presence of additional surface features. However, it may also indicate that the surface of Vega is not stable on decade timescales, in which case we need to apply a model that can better address its time variability. 

Quasi-periodic Gaussian processes \citep{Rasmussen:2006,Roberts:2012} are known to represent the effects of active regions rotating in and out of view \citep{Haywood:2014, Rajpaul:2015, Angus:2018}. Using {\tt radvel}, we model the correlated stellar noise with a kernel matrix whose elements are defined as
\begin{equation}
    C_{ij} = \eta_1^2 \exp{\left(-\frac{\left|t_i-t_j\right|}{\eta_2^2} - 
    \frac{\sin{\left(\frac{\pi \left|t_i-t_j\right|}{\eta_3}\right)}}{2\eta_4^2}\right)},
\end{equation}
where $t_i$ and $t_j$ are observations made at any two times; $\eta_1$ is the amplitude hyperparamter; $\eta_2$ is the exponential decay timescale, which is physically related to the lifetime of the active regions; $\eta_3$ is the period of the variability, corresponding to the rotational period; and $\eta_4$ is the length scale of the periodic component, describing the high-frequency variation in the stellar rotation. It is important to note, however, that these physical interpretations are often not straightforward, particularly given the degeneracies between the hyperparameters.

\begin{deluxetable}{llc}
\tabletypesize{\small}
\tablecaption{Stellar activity RV models\label{tab:activity}}
\tablehead{\colhead{Parameter~~~~~~~~~~} & 
           \colhead{Units~~~~~~~~~~} & 
           \colhead{Value}}
\startdata
\smallskip\\[-3.5ex]
%%%%%%%%%%%%%%%%%%%%%%%%%%%%%%%%%%%%%%%%%%%%%%%%%%
\multicolumn{2}{l}{Multiple Keplerians} & \\
~~~$P_1$      & days & $0.676682 \pm 0.000006$ \\
~~~$T_{c1}$   & BJD  & $2457187.123 \pm 0.013$ \\
~~~$e_1$      & \nodata & $0.438 \pm 0.052$ \\
~~~$\omega_1$ & deg  & $344 \pm 12$ \\
~~~$K_1$      & \ms  & $11.6 \pm 1.2$ \\
~~~$P_2$      & days & $0.676086 \pm 0.000006$ \\
~~~$T_{c2}$   & BJD  & $2457187.195 \pm 0.051$ \\
~~~$e_2$      & \nodata & $0.46 \pm 0.22$ \\
~~~$\omega_2$ & deg  & $179 \pm 15$ \\
~~~$K_2$      & \ms  & $11.1 \pm 3.8$ \\
~~~$\gamma$   & \ms  & $-33.69 \pm 0.69$ \\
~~~$\sigma_{\rm jit}$   & \ms  & $13.1 \pm 1.0$ \\[0.8ex]
\hline\\[-2.5ex]
%%%%%%%%%%%%%%%%%%%%%%%%%%%%%%%%%%%%%%%%%%%%%%%%%%
\multicolumn{2}{l}{Quasi-Periodic Gaussian Process} & \\
~~~$\eta_1$~~~~~~~~~~~~~~~~~~~~~~~ & \ms  & $14.4^{+2.2}_{-1.8}$ \\
~~~$\eta_2$ & days & $179^{+65}_{-58}$ \\
~~~$\eta_3$ & days & $0.6764^{+0.13}_{-0.00021}$ \\
~~~$\eta_4$ & \nodata & $0.44^{+0.16}_{-0.15}$  \\
~~~$\gamma$ & \ms  & $-33.9^{+2.9}_{-2.8}$ \\
~~~$\sigma_{\rm jit}$ & \ms  & $12.9 \pm 1.4$ \\[0.8ex]
\enddata
\end{deluxetable}

No trend is apparent in our data or the residuals of early fits and we choose to exclude a linear or quadratic term in our model. However, we do include offset and jitter terms to account for instrumental offset and noise. Priors are only placed on parameters to keep them within physically possible limits. The complete log-likelihood of this model is
\begin{equation}
    \ln \mathcal{L} = -\frac{1}{2}r^TK^{-1}r - \frac{1}{2}\ln{\left(\det K\right)} - \frac{n}{2}\ln{\left(2\pi\right)},
\end{equation}
where $r$ is the vector of residuals, $K$ is the covariance matrix, and $n$ is the number of data points.

We perform an affine-invariant Markov chain Monte Carlo (MCMC) exploration of the parameter space using the ensemble sampler {\tt emcee} \citep{ForemanMackey:2012,DFM:2019}. Our MCMC analysis used 8 ensembles of 50 walkers and converged after $4950000$ steps, achieving a maximum Gelman-Rubin statistic \citep{Gelman:1992} of $1.006$. The resulting posterior distributions are shown in \rftabl{activity}. We do note that this model is over-fitting the data, with a reduced $\chi^2$ statistic of $0.869$; either the model is fitting noise or our cadence is not sufficient to constrain high-frequency variations in the rotation signal. In either case, the Gaussian process is likely too complex of model for our data. We check our results with a second GP fit of the radial velocities using a quasi-periodic kernel with the {\tt PyMC3} Python package \citep{Salvatier:2016}, which returns results well within $1\sigma$ of those in \rftabl{activity}.

\subsection{Additional Signals of Interest}

The activity signal at the period of stellar rotation is by far the strongest we observe, but other candidate signals have been reported in previous Vega data sets, and we also search for signals of lower significance in the TRES RVs.

\subsubsection{Previously Suggested Signals}

\cite{Bohm2015} used their SOPHIE spectra to search for additional short-period signals and reported a possible detection at $1.77\ {\rm d}^{-1}$\ ($0.56$\ days) with an amplitude of $6$\ \ms. If it has a planetary origin, it would correspond to roughly a Saturn-mass planet well aligned with the stellar spin near the co-rotation radius of the star. However, no such periodicity appears in our TRES observations. When we inject this signal into our data, we find that it would have been clearly detected, which suggests that the signal in the SOPHIE data set does not correspond to a planet. While we cannot conclude with certainty what the source of that signal was, the SOPHIE data only span $5$\ nights and cover about $7$\ hours each night, so one possibility is that it may have resulted from a combination of short-timescale stellar variability and the sampling.

\subsubsection{Signals at Long Periods}

Because our Gaussian process model of the $0.677$-day signal over-fits the data, it is difficult to use the residuals of that model to search for---or to jointly fit---additional signals. However, we note that in the original periodogram of our radial velocities, the strongest peak with a timescale longer than a few days is located at $197.3 \pm 4.8$\ days. This signal is not formally significant, though a Keplerian fit converges (to $194$\ days) and implies a  minimum mass similar to that of Saturn. It is possible that the presence of the high frequency activity signal is suppressing its significance and that careful modeling of the activity may enhance it. We offer several possible explanations for this peak and investigate the effect of stellar activity modeling. 

The first explanation is that it simply arises from white noise. Since it is not formally significant, it would not be unusual to observe it by chance. A second, similar interpretation is that the roughly half-year period may be an artifact of the observing cadence, given that our observations are necessarily seasonal, with breaks at the same time each year when Vega is behind the Sun and also during the August monsoon season in Arizona. There is indeed a small peak in the window function near the same period ($182.8 \pm 3.5$\ days; \rffigl{periodogram}). Our RV standard stars, which have been observed with similar cadence for similar time spans, show similar window functions but do not show similar periodogram peaks, which seems to suggest the window function is not to blame. However, it is possible that Vega could be susceptible to systematics not seen in the standard stars, which have exposure times an order of magnitude longer. For example, in the uncorrected standard star RVs, we observe a slight seasonal periodicity that is corrected in our zero point analysis. If the tracking during very short exposures of Vega is imperfect, this could lead to (systematic) differences in the illumination of the fiber as a function of position on the sky, and ultimately leave uncorrected instrumental effects that vary with season. As a result, we cannot rule out instrumental and window function effects as the source of this signal.

A third possibility is that although the timescales are very different, the signals at $0.677$\ days and $194$\ days may both originate from activity; the timescale of evolution of the former ($179^{+65}_{-58}$\ days) derived from the GP fit is consistent with the latter. 
Because the GP activity model is flexible enough to absorb the long-period signal even if it is real, we examine the effect of the multiple-Keplerian model on this signal. If active regions are long-lived and exist with very slightly different periods due to differential rotation, then beating between the two frequencies will lead to evolution of the observed variation on long timescales. Modeling and removing the signal from the static active regions should reduce the significance of signals related to the beat frequency, but should not generally absorb unrelated signals like the GP does. After fitting the two dominant activity signals near $0.677$\ days, the FAP of the peak at $194$\ days is reduced (to $\mysim 5\%$). However, fitting one more Keplerian corresponding to the next most significant harmonic ($P_{\rm rot}/2$) removes the 194-day signal completely. 

Given the very modest significance of the $194$-day signal, the existence of plausible alternative explanations, and its disappearance when removing Keplerians associated with the rotation period, we do not consider this to be a planetary candidate. There are no other long-period signals of note in our RVs.

\begin{deluxetable}{llcc}
\tabletypesize{\footnotesize}
\tablecaption{Candidate Planetary Companion to Vega~~~~~~\label{tab:planet}}
\tablehead{\colhead{Parameter} & 
           \colhead{Units} & 
           \twocolhead{Value~~~~~~~~~~~~~~~~~~}}
\startdata
\smallskip\\[-3.5ex]
                   &         & Eccentric             & Circular              \\
$P$             & days    & $2.42977 \pm 0.00016$   & $2.42971 \pm 0.00018$   \\
$T_c$           & BJD     & $2457186.51 \pm 0.18$   & $2457186.631 \pm 0.067$ \\
$e$             & \nodata       & $0.25 \pm 0.15$         & $0$                     \\
$\omega$        & deg     & $304 \pm 40$            & \nodata               \\
$K$             & \ms     & $6.4 \pm 1.1$           & $6.1 \pm 1.1$           \\
$\gamma$        & \ms     & $-34.30 \pm 0.75$     & $-34.34 \pm 0.75$     \\
$\sigma$        & \ms     & $9.7 \pm 1.0$         & $9.7 \pm 1.0$         \\
\hline\\[-2.5ex]
$m_p\sin{i}$    & \mearth & $21.9 \pm 5.1$          & $21.2 \pm 3.8$          \\[0.8ex]
\enddata
\end{deluxetable}

\subsubsection{A Candidate Planetary Companion}

\begin{figure}[!t]
\centering\includegraphics[width=\linewidth]{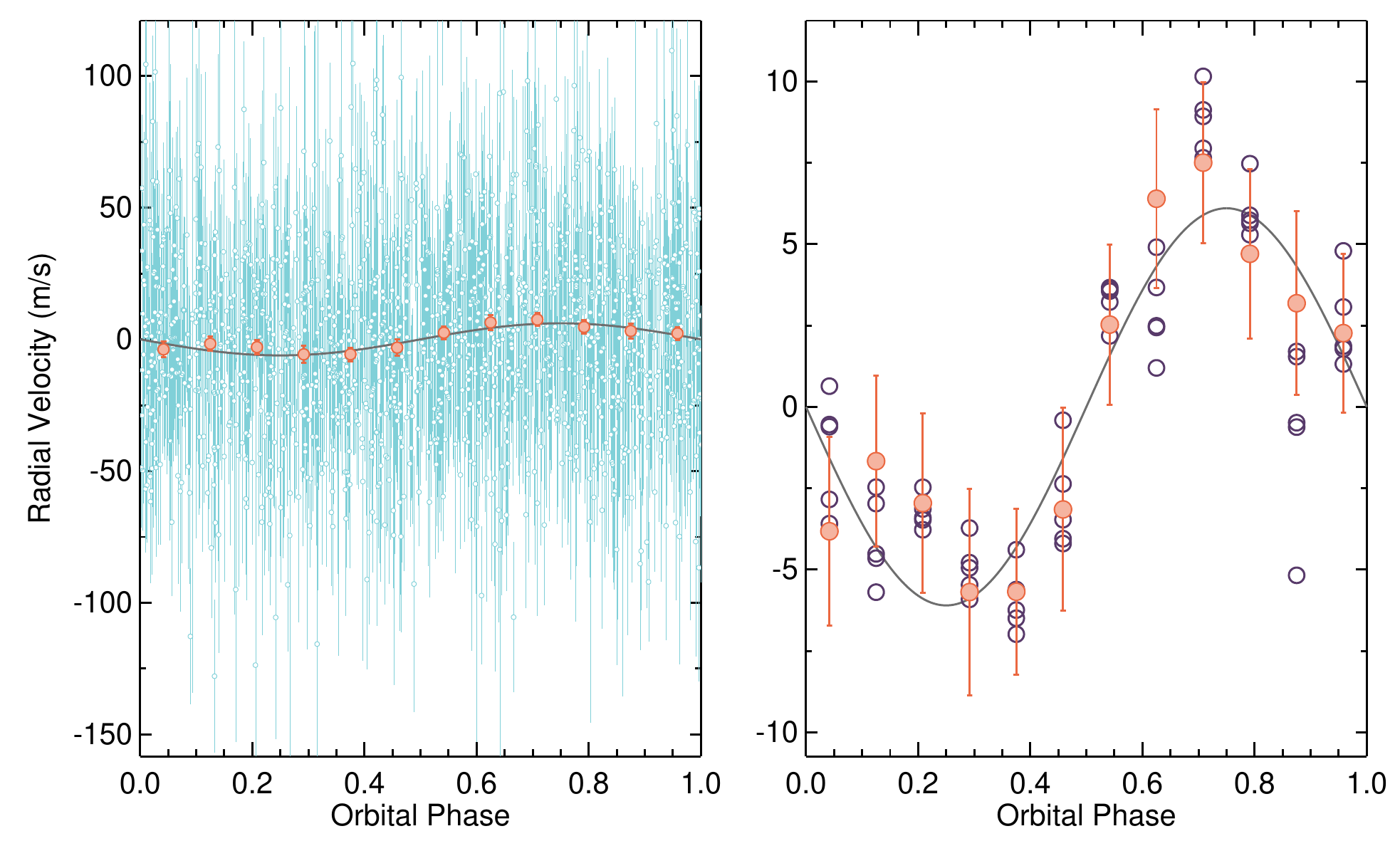}\\[3ex]
\caption{\textit{Left}: TRES RVs (small light blue circles) phase-folded to the $2.43$-day period of a candidate planet orbiting Vega after removal of activity signal by fitting Keplerians at the rotation period. Data binned in phase are shown as orange circles. \textit{Right}: The same orbit on a different velocity scale to better show the binned data. The purple open circles represent phased, binned data from the intermediate steps of whitening the RVs by removing Keplerians associated with the rotation period. A consistent orbit is seen when removing 0, 1, 2, 3, 4, or 5 activity signals.}
\label{fig:planet}
\end{figure}

Interestingly, after removing two Keplerian signals associated with the stellar rotation, the strongest remaining peak has a period of $2.43$\ days and a formal FAP of only $<0.01$. Unlike the $194$-day signal, it is robust to the inclusion of additional Keplerians near the rotation period and its harmonics; the FAP of the 2.43-day signal remains below $1\%$\ whether we account for stellar noise or not, and even if we include more complex static models of the stellar surface with as many as 5 Keplerian components. We do not find any relationship between this period and the rotation period and its aliases listed in \rftabl{signals}, and the uncertainties on the periods are much smaller than the differences between them. A circular orbit with a period of $2.43$\ days has an RV semi-amplitude of about $6$\ \ms, corresponding to a minimum mass of about $20$\ \mearth. According to the difference in the Bayesian Information Criterion ($\Delta{\rm BIC}$), a circular orbit is statistically preferred, but we present both eccentric and circular solutions in \rftabl{planet}. \rffigl{planet} shows the phase folded RVs of the candidate planetary signal. While we cannot conclusively rule out false positive scenarios, we discuss in \rfsecl{discussion} ways in which we might confirm the candidate with future observations and analyses.

\section{Detection limits}

\subsection{TRES Radial Velocities}

\subsubsection{Isotropic Orbits}
\label{sec:iso-aligned}

We begin the calculation of our detection limits by randomly generating $\mysim 10^6$ models corresponding to planets with semi-major axes between $0$ and $15$\ au and masses ranging from $0$ to $100$ \mjup. Each orbit is assigned an inclination (drawn from a uniform distribution in $\cos{i}$), eccentricity \citep[drawn from a beta distribution described by][with parameters $a=0.867$ and $b=3.03$]{Kipping:2013}, an argument of periastron (drawn from a uniform distribution), and a time of periastron passage (drawn from a uniform distribution). With {\tt radvel}, we calculate the expected radial velocities for each orbit at our times of observation. We then add noise scaled to the observed uncertainties at each time stamp. Using {\tt radvel}, we fit a flat line to the synthesized RVs and calculate the $\chi^2$ statistic and its associated $p$ value for this model. Low $p$ values indicate that the synthesized signal is unlikely to arise from white noise---i.e., it is distinguishable from a flat line and we therefore consider it detected. To set a threshold for detection, we follow \citet{Latham:2002}, who demonstrate that $p < 0.001$ is a conservative threshold below which signals are almost entirely real. This makes sense, since data comprising only white noise will exhibit $p<0.001$\ only $0.1\%$\ of the time. While we find that other methods of injection recovery---such as a requirement that $\Delta{\rm BIC}$\ between the detected signal and a flat line model be greater than $10$---yield more sensitive limits, we adopt the conservative $p(\chi^2)<0.001$\ because we do not address correlated noise in our simulations; in the likely event that we cannot perfectly model the stellar activity, it may hinder our ability to detect some signals. The top panel of \rffigl{detection} shows the distribution of detection probabilities for isotropically oriented orbits, from which it can be seen that we are sensitive to sub-Saturn masses orbiting at $0.1$\ au and masses as low as about $2\ \mjup$\ at $6$\ au, which corresponds to an orbital period of $10$\ years around Vega. Beyond $6$\ au, our detection limits degrade more quickly, as our data no longer cover an entire orbit. Nonetheless, we are sensitive to the most massive giant planets all the way out to $15$\ au. Given the roughly isotropic distribution of stellar obliquities for hot stars hosting transiting hot Jupiters, the decision to draw inclinations from an isotropic distribution may be the most realistic assumption for short periods. However, it is unclear whether the same is true at long periods. We therefore also explore our detection limits for well-aligned orbits in the following section.

\subsubsection{Well-Aligned Orbits}
\label{sec:well-aligned}

\begin{figure}[!htb]
\centering\includegraphics[width=\linewidth,trim={.45cm .15cm .45cm .45cm},clip]{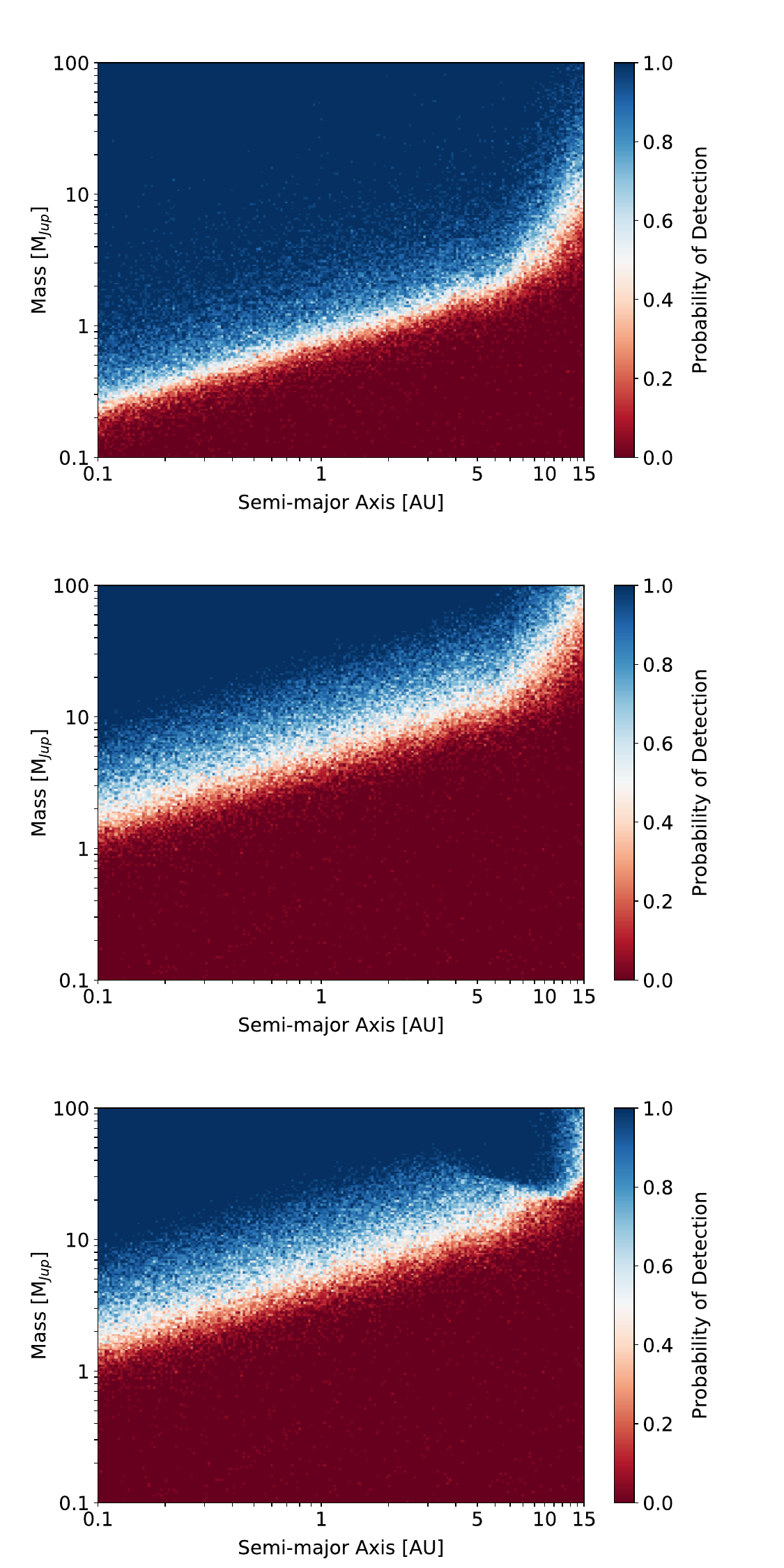}
\caption{Top: TRES detection probabilities for objects in isotropically distributed orbits. Middle: Detection probabilities for objects in orbits well-aligned with Vega's spin-axis. Bottom: TRES detection probabilities combined with Palomar P1640 detection limits from \citet{Meshkat2018} for objects in well-aligned orbits.}
\label{fig:detection}
\end{figure}

For our purposes, we will consider a well-aligned orbit to fall within $5\degree$ of the stellar spin axis. Assuming Vega has an inclination of $6.5 \degree$, this means that a well aligned planet would have an inclination between $1.5\degree$ and $11.5\degree$. To calculate detection limits for well-aligned orbits, we follow the same steps as in the previous section, but instead assign inclinations drawn from a uniform distribution between $1.5\degree$ and $11.5\degree$. The resulting detection probabilities are shown in the middle panel of \rffigl{detection}. While drawing from a distribution of well aligned orbits (highly inclined to the line of sight) clearly reduces detection probabilities, we are still sensitive to the most massive giant planets as widely separated as $6$\ au; beyond this distance, we are only sensitive to brown dwarfs or stars. 

\subsubsection{Including Direct Imaging Limits}

The direct imaging limits from \cite{Meshkat2018} also explore the inner $15$\ au around Vega for planets. While our RV detection limits fall off exterior to $\mysim 6$\ au, \cite{Meshkat2018} have limited sensitivities \textit{interior} to this boundary. We also note that direct imaging and radial velocities are most sensitive to planets at opposite inclinations. Consequently, a brown dwarf that is missed by RVs because it is highly inclined to the line of sight may be detected by direct imaging; one that is too close in projection for direct imaging can be detected by RVs. By combining our results, we are able to provide a more comprehensive limit on the presence of widely separated companions. 

We calculate the projected separation for each randomly drawn sample in \rfsecs{iso-aligned}{well-aligned} on UT 2016 Aug 19 and 2017 June 05 (two of the observing times in \citealt{Meshkat2018}) using Equation (7) in \cite{Kane:2011}. We consider the planet to be detectable in the data presented by \cite{Meshkat2018} if its mass falls above the five-sigma H-band mass limits for either of the observation times. We then use the same criterion as before to determine if it is detectable in our radial velocities. For isotropically distributed orbits, our RVs are more sensitive than the imaging at all separations, but for long-period orbits aligned with the stellar spin, imaging is more sensitive. The detection probabilities for well-aligned orbits using RVs and direct imaging are shown in the bottom panel of \rffigl{detection}.

\subsection{TESS Photometry}
\label{sec:injection}

Using the Python package {\tt batman} \citep{Kreidberg:2015}, we randomly generate $\mysim10^5$ transit models for planets with periods ranging from 0.5 to 30 days and radii between 1 and 8 \rearth. Each orbit is assumed to be circular and semi-major axes are calculated using Kepler's law, assuming that the planet's mass is negligible compared to Vega. Transit times are randomly assigned from a uniform distribution and inclinations are drawn from a uniform distribution in $\cos{i}$ such that
\begin{equation}
    0 < \cos{i} < \frac{R_{\star} - R_p}{a},
\end{equation}
where $a$ is the semi-major axis and $R_\star$ and $R_p$ are respectively the stellar and planetary radii, ensuring that the planet transits. Additionally, each synthetic transit follows quadratic limb darkening laws for an A0 star in the \tess\ bandpass with $u_1 = 0.1554$ and $u_2 = 0.2537$ \citep{Claret:2018}.

We then inject each model into our light curve and conduct a transit search using {\tt TLS}. We consider a transit to be detectable if the best period recovered by {\tt TLS} is within $1\%$ of the injected period and if at least one of the transit times is within the same margin of an injected transit. Additionally, because a transit could appear within the gaps of our light curve, any recovered integer multiples of the injected period are considered detectable.

We also require a transit to be distinguishable from a false positive to be considered a true recovery, ensuring that the signal would not be dismissed due to low signal in a real search. Using synthetic light curves and transits, \cite{Hippke:2019} find that a signal detection efficiency (SDE) threshold of 7 corresponds to a false positive rate of $1\%$ for {\tt TLS}. Therefore, for a transit to be detectable, we also require that the best period have an SDE of 7 or greater. 

\begin{figure}[!ht]
\centering\includegraphics[width=1.08\linewidth]{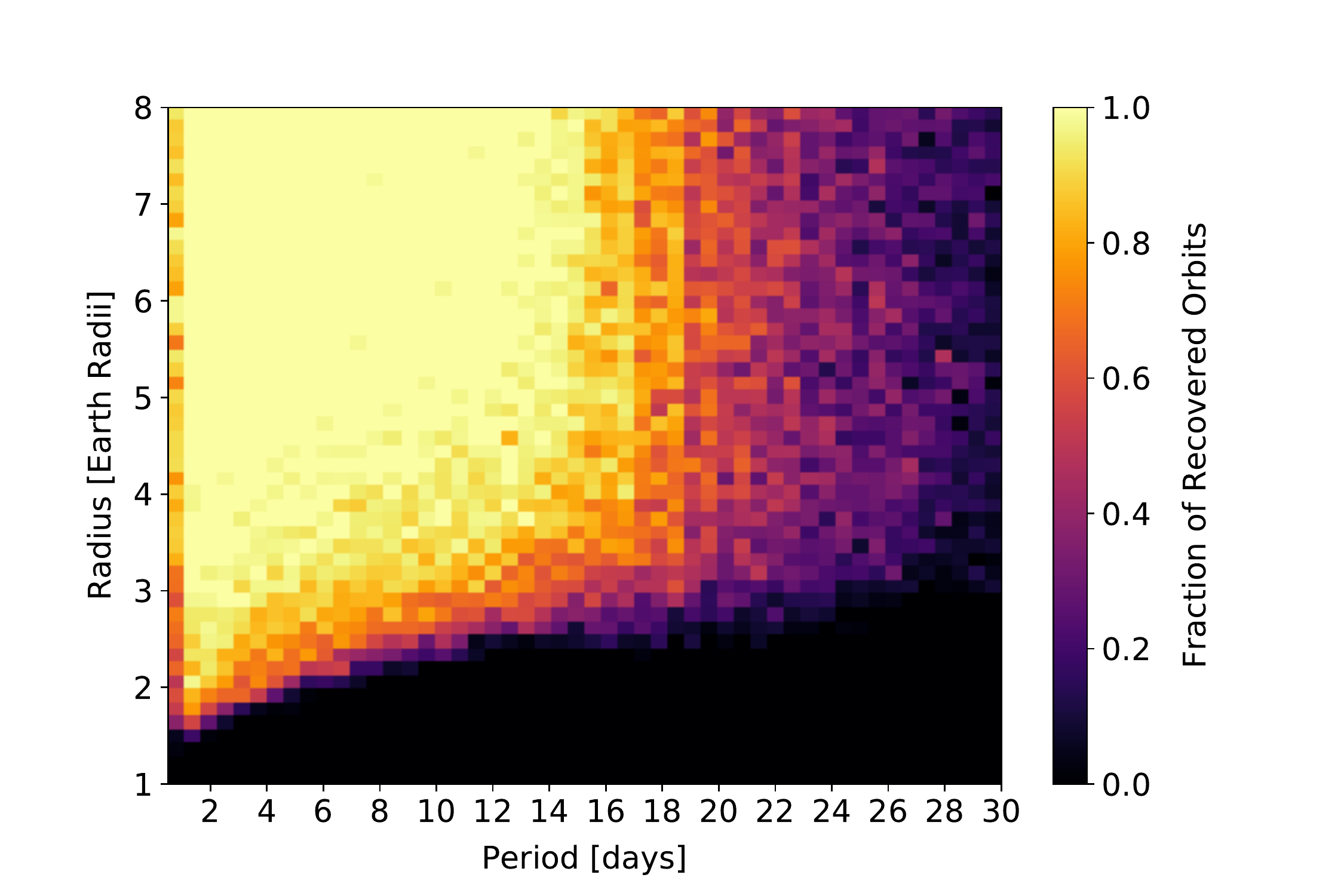}
\caption{Transit detection limits for \tess\ observations of Vega. For periods shorter than a few days, we can rule out most transiting planets with radii greater than 2 \rearth, and for periods out to 15 days, planets with radii greater than 3 \rearth. Planets with longer periods do not transit frequently enough to be consistently ruled out. However, we note that for many of the larger injected planets not recovered according to our criteria, single transit events would be easily visible by eye, so our detection map should be viewed as a conservative limit.}
\label{fig:transit}
\end{figure}

The results of our injection recovery test are shown in \rffigl{transit}. At the shortest periods, we are sensitive to transiting planets as small as $2$\ \rearth, and as small as $3$\ \rearth\ for orbital periods similar to the duration of a \tess\ orbit ($\lesssim 14$\ days). For periods longer than a couple weeks, our formal sensitivity drops, as fewer transits are expected; some planets may exhibit only one or two transits total, and some might not even be observed once by \tess\ due to data gaps between orbits. Many of these will not pass our automated TLS criteria for detection. On the other hand, the transits of even relatively small planets orbiting Vega would be easily visible by eye in the quiet \tess\ light curve. A $4$ \rearth\ transit, for example, would be about 180 ppm. It is clear that there are no Neptune-sized planets transiting in the \tess\ data even once. While the formal detection limits shown in \rffigl{transit} are illustrative of the types of transiting planets we could detect most easily, they should be taken as a conservative estimate.

\section{Discussion}
\label{sec:discussion}

\subsection{Stellar Activity}

\begin{figure*}[!htb]
\centering\includegraphics[width=\linewidth]{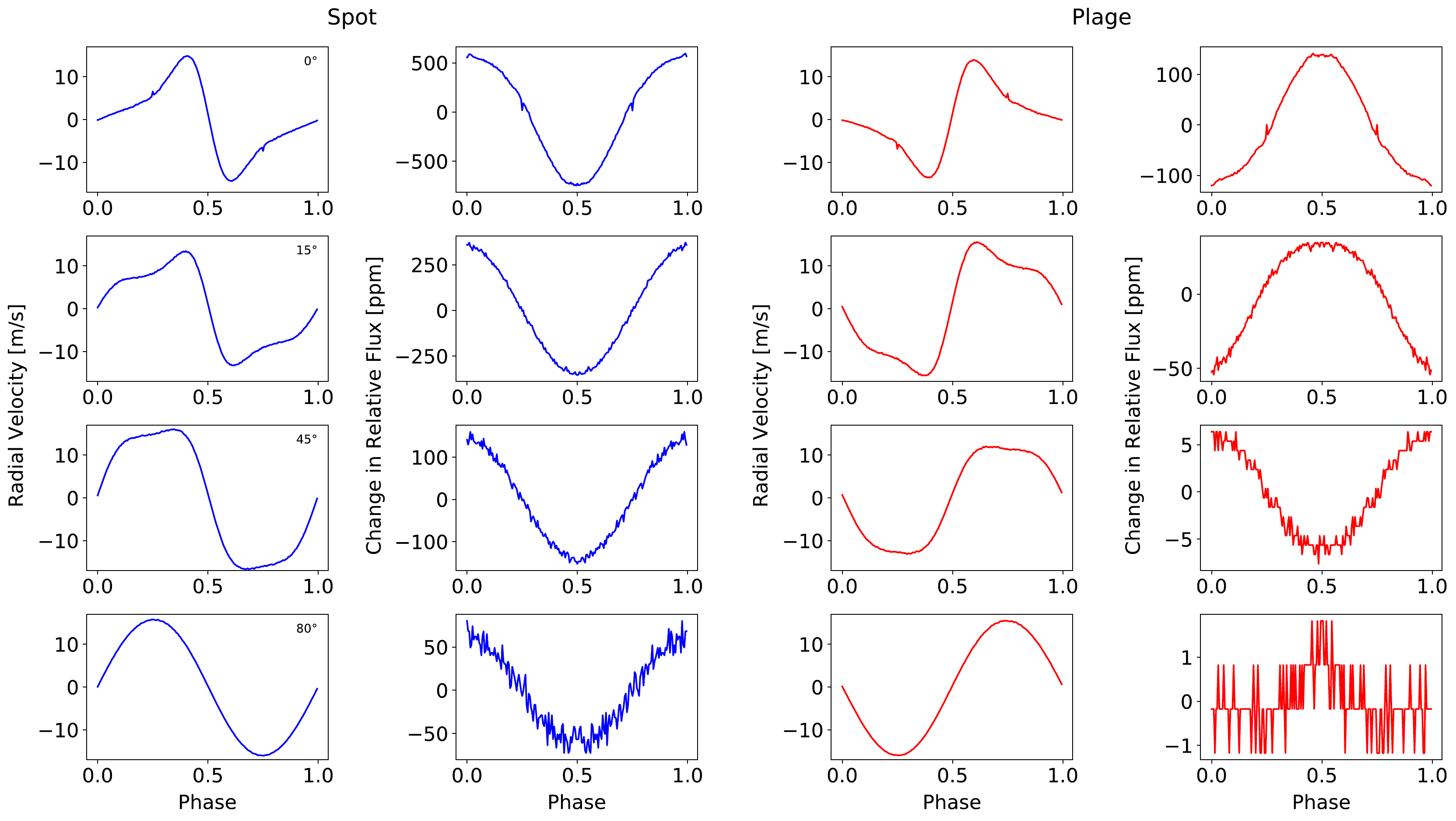}
\caption{{\tt SOAP 2.0} simulations of the radial velocity and photometric variation induced by spots (blue) and plages (red) at different latitudes. The top row simulates a spot or plage on Vega's equator, the second row a latitude of $15\degree$, the third a latitude of $45\degree$, and the last a latitude of $80\degree$. Dark spots deviate from Vega's effective temperature by 750 K and the sizes of active regions are varied to obtain an RV amplitude similar to the observed values of $\mysim 15$ ms$^{-1}$. While spots and plages are both able to reproduce the RV variation, the change in flux is much lower for plages. Vega's surface may be more complicated than this toy model, but this suggests that plages are more likely the dominant driver of the radial velocity signal and the quiet \textit{TESS} light curve.}
\label{fig:soap}
\end{figure*}

Rotational modulation has consistently been found in spectroscopic observations of Vega \citep{Bohm2015, Petit:2017}, providing evidence for active regions on the surface of the star. While chemically peculiar A stars are known to have star spots, normal A stars conventionally do not exhibit the same traits. Vega is the first normal A-type star observed to have a weak surface magnetic field \citep{Lignieres:2009, Petit2010}, accounting for the rotational modulation. However, the origins of these magnetic fields are uncertain. One possible mechanism is dynamo action, where thin convective layers host a dynamo driven by convective motion, generating a magnetic field. On the other hand, the fields could be `failed fossils,' which are generated in the early phases of a star's life and dynamically evolve towards fossil equilibrium \citep[e.g.,][]{braithwaite:2013}. These models can be distinguished by the time variation of the magnetic field. Dynamos are intrinsically variable, leading to spot lifetimes similar to the rotation period. In contrast, failed fossil fields would evolve much more slowly.

Vega appears to have a very complex magnetic field. \cite{Bohm2015} find no evidence for rapid variability consistent with dynamo action in their data, suggesting the presence of star spots that last for over five days. However, applying Doppler imaging techniques to the same data set, \cite{Petit:2017} do find rapidly evolving surface features in combination with stable structures. \cite{Cantiello:2019} suggest that this rapid variation is a sign of dynamo-generated magnetic fields. 

While our spectroscopic observations are likely not high-cadence enough to characterize the short term evolution of co-rotating structures, they stretch over a span of time much longer than the five consecutive nights used by \cite{Bohm2015} and \cite{Petit:2017}, which may provide insight regarding the long-term evolution. In \rfsecl{rvsignals}, we described that our RVs can be modeled by either a Gaussian process with a quasi-periodic kernel or the combination of Keplerian signals very close to the rotation period. In both cases, the implied timescale for evolution is long: the GP fit suggests an exponential decay timescale of half a year, while the Keplerian model that adequately reproduces the signal seen in our ten-year data set suggests that some surface features may be even more long-lived. It is hard to distinguish between the two models, but both imply the presence of surface features with lifetimes longer than expected for dynamo fields. Even so, this is not necessarily at odds with the complex and time-varying reconstructed surface map presented by \citet{Petit:2017}, who identified both variable and stable surface features in their data. We speculate that their stable features may be the ones responsible for our long-lived RV signal, while their rapidly varying features contribute high frequency noise to our data that we cannot characterize due to our limited observing cadence. We do, however, measure stellar jitter with a magnitude comparable to that of the coherent modulation, which may arise from rapidly varying surface features. If both types of features exist, it may imply magnetic structures influenced both by a failed fossil field and a subsurface dynamo. However, long term spectropolarimetric observations of Vega are necessary to conclusively determine the behavior of its active regions.

Another interesting feature of Vega's activity is its low photometric variation, particularly given its maximum radial velocity amplitude of $14.4^{+2.2}_{-1.8}$\ \ms. A dark spot inducing a $10$\ \ms\ variation on a star rotating with $\vsini=20$\ \kms\ should also induce a photometric signal of $A_{\rm RV}/\vsini\mysim500$\ ppm \citep{Vanderburg:2016e}; the observed amplitude of only $10$\ ppm provides independent evidence that dark spots do not dominate the surface. To explore this further, we use {\tt SOAP 2.0} \citep{Dumusque:2015} to compare the effects of star spots and plages at different latitudes on a Vega-like star (\rffigl{soap}). We vary the size of the active region to reproduce an RV signal with an amplitude of $\mysim 15$\ \ms\ and compare the expected photometric signals. We find that a dark spot would induce photometric variations ranging from a hundred to over 700 ppm. On the other hand, a high-latitude plage can reproduce photometric variations less than 10 ppm, consistent with the observed \tess\ light curve. This is because plages are only marginally hotter than the rest of the stellar surface, resulting in little flux variation, but are still surrounded by local magnetic fields that suppress convective blueshift and cause regions to appear redshifted, creating an RV signal \citep{Dumusque:2015}. Given that \citet{Petit:2017} detected many bright and dark regions on Vega's surface, the picture is clearly not as simple as our simulations, but it suggests that the observed variations are primarily driven by plages.

\subsection{Candidate Planets Orbiting Vega}

Any planetary signals in our radial velocities may be difficult to detect, since the dominant signal arises from stellar activity and its removal is not straightforward. Nevertheless, we are able to rule out the presence of a candidate signal near 0.53 or 0.56 days previously reported by \cite{Bohm2015}, and we do identify two signals worthy of further investigation. The first, with a period of $196.4^{+1.6}_{-1.9}$\ days, would have a semi-major axis of $0.853^{+0.020}_{-0.022}$\ au and a minimum mass of $0.252\pm0.066$\ \mjup, falling well below the detection limits of previous direct imaging surveys. The power of the signal monotonically increases with the addition of new data, which is what one would expect for a real orbiting companion. However, it is not formally significant, is located near a small peak in the window function, and its significance is reduced further when the activity signal is modeled. We conclude that there is not good evidence for a planet at this period. 

The second interesting signal emerged from the velocity residuals when modeling the activity with multiple Keplerians near the rotation period. We identify a short-period signal with a formal false alarm probability less than $1\%$, a best-fit period of $2.43$\ days, and a semi-amplitude of $6$\ \ms, implying a minimum mass of $\mysim 20$\,\mearth. A true mass this low would require a polar orbit, but A stars hosting short-period planets do display a wide range of stellar obliquities, and even a highly inclined orbit that is well aligned with the stellar spin ($i=6.2$\,deg) would correspond to a planetary mass companion ($\mysim 0.6$\,\mjup). 
Continued radial velocity observations could provide further insight into the presence of a planet orbiting Vega, but more work needs to be done to account for the activity signal; further spectroscopic observations should be planned carefully, for example to achieve a cadence that resolves the rotational and orbital periods, allowing better simultaneous modeling of planet and stellar activity.

If there is a planet with a period of $2.43$\ days orbiting Vega, there are a few ways one might confirm its existence. \tess\ data rule out transits of objects larger than about $2$\ \rearth\ at this period, so unless it is very dense, this candidate planet does not transit. It has taken 1500 spectra obtained over 10 years to detect the RV signal at low significance, so it may require a large investment of spectroscopic resources to increase the SNR of the detection even if it is real. A high cadence campaign spanning many orbital periods (ideally from multiple longitudes to achieve uninterrupted data) might be best suited for mitigating stellar activity and limiting its evolution during observations. Alternatively, it may be possible to directly detect the planet using high resolution spectroscopy and cross-correlation against a template of elemental or molecular species in the planetary atmospheres \citep[e.g.,][]{snellen:2010}. With such a short period around such a hot star, the planet would have an equilibrium temperature of $\mysim3250$\ K (assuming a Bond albedo of 0.25), and would be the second hottest known exoplanet after KELT-9b \citep{Gaudi2017}. KELT-9b and several other hot Jupiters have had atmospheric lines successfully detected with high resolution spectroscopy \citep[e.g.,][]{brogi:2012,birkby:2013,yan:2019}. Importantly, these detections are not reliant on a transiting geometry. To predict the expected signal, we need to estimate the star-to-planet contrast ratio, and therefore the planetary radius. Nearly all highly irradiated giant planets have inflated radii (some as large as $\mysim2$\ \rjup), so $1$\ \rjup\ would be a conservative estimate.
As outlined in \citet{birkby:2018}, the signal to noise of the planetary signal is described by 
\begin{equation}
\rm SNR_{planet}=\left(\frac{S_{p}}{S_{\star}}\right) SNR_{star} \sqrt{N_{lines}},
\label{eqn:SNR}
\end{equation}
where $S_{p}/S_{\star}$\ is the planet-to-star contrast ratio. Adopting $R_p = 1$\ \rj\ and assuming blackbody radiation, the planet-to-star contrast ratio for the candidate Vega planet would be $2 \times 10^{-5}$. As a reference, assuming $1$\ \rj\ for tau Bo\"otis b \citep[the first non-transiting planet with a detection using this technique;][]{brogi:2012}, we would estimate a contrast of $2.6 \times 10^{-5}$. The total SNR of our Vega spectra is about 26,000, implying ${\rm SNR_{planet}} \mysim 0.5 \sqrt{\rm N_{lines}}$. If we can resolve $\mysim 100$\ lines in the planetary spectrum, it is possible that we could detect such a planet with our TRES data. However, a high confidence detection may have to wait for a data set better suited to this purpose. The SOPHIE data obtained by \citet{Bohm2015} have higher resolving power and total SNR than our TRES data but cover a narrower wavelength range. TRES covers the red optical where contrasts are not as extreme and where lines from molecules such as TiO, CO, and ${\rm H_2O}$\ may be present in the planetary atmosphere. On the other hand, \ion{Fe}{1} has been detected on the daysides of ultra-hot Jupiters like KELT-9b and WASP-33b, so direct detection of the planet is possible even in the blue optical. Ultimately, it may be necessary to move to the near infrared where contrasts are further reduced and the presence of strong molecular bands of CO and ${\rm H_2O}$\ can boost the SNR.

\subsection{Current and Future Limits on Planetary Companions}
Using 1524 TRES RVs, we are able to place new detection limits on planets orbiting within 15 au of Vega, a region in which direct imaging surveys have only ruled out brown dwarfs. Assuming no preference for the orientation of the orbital plane---consistent with the observed distribution for short-period planets orbiting hot stars---we can rule out nearly all hot Jupiters. We are sensitive to Saturn-mass objects at 0.1 au, Jupiters at 1 au, 5 \mjup\ at 10 au, and 13 \mjup\ at 15 au. For a planet well aligned with the stellar spin, these masses increase by a factor of about 8 and even massive planets would therefore be difficult to detect beyond $\mysim 3$\ au. Nevertheless, these limits indicate that brown dwarf companions in any orientation are exceedingly unlikely within 3 au. While the majority of planets that have been hypothesized from the observed architecture of Vega's disk would reside beyond 15 au, \cite{Raymond2014} find that systems of planets with masses ranging from $\mysim1$\ \mjup\ at $5$--$10$\ au to Neptune masses at tens of au could replenish the hot dust in the inner belt. Our data are not quite sensitive enough to constrain planets this small at these distances, with our detection probabilities falling off steeply below about $2$\ \mjup\ at 5 au, even for planets drawn from an isotropic inclination distribution.

With \textit{TESS} data, we are able to place constraints on the size of any transiting planets. While most planets would not transit, requiring both a very high stellar obliquity and a semi-major axis $\lesssim 0.2$\ au, the extremely precise photometry obtained by \tess\ results in very sensitive detection limits, despite the large stellar radius. We can rule out any transiting hot Jupiters, along with most other planets with radii greater than $3$\ \rearth. Further observations could place even more extensive limits on transiting planets while helping us to better understand the active regions on the star. The \textit{TESS} extended mission will return to the northern hemisphere in its second year of operations, which may be the next opportunity for additional uninterrupted measurements at a similar precision. The 10-minute full frame images used in the extended mission will also improve the time resolution by a factor of three, opening the possibility of further characterization of high frequency stellar variations.

Current and future space telescopes offer additional opportunities to search for planets around Vega. \cite{Meshkat2018} show that planned James Webb Space Telescope (JWST) NIRCam GTO observations of Vega \citep{Beichman:2010} will have greater sensitivity than previous surveys, perhaps extending to Saturn-mass planets. Additionally, MIRI GTO observations are expected to resolve the potential asteroid belt analog \citep{Beichman:2017}, providing further insight into the disk structure of Vega and its implications for a planetary system. However, JWST will only be able to search the region beyond $1.5$\arcsec\ (11 au) from Vega for planets; any closer, and the star will saturate the instrument. The Nancy Grace Roman Space Telescope coronagraph instrument (CGI) is more promising for close companions, as it is intended to observe small fields around bright stars. Because Vega is so bright, Roman would not need to observe a bright PSF reference star, potentially allowing better CGI stability and improving contrast for point sources. On the other hand, the CGI is optimized for stars with angular diameters less than $\mysim 2$\ mas; with an angular diameter over $3$\ mas, there may be some light leakage when observing Vega. While it is uncertain how these factors would balance out, if the conventional Roman CGI sensitivity limits apply to the star, the mission could observe Jupiter-sized planets within $0.2$\arcsec\ (close to $1$\ au) \citep{Nemati:2017, Krist:2018}. Combining Roman and JWST observations, future space missions promise to place new direct imaging limits on planets throughout the entire Vega system. 

As a nearby star, Vega is also an interesting candidate for astrometric detection of companions, though the best precision---i.e., with Gaia---may not be possible. \cite{Sahlman:2016,Sahlmann:2018} describe protocols to observe very bright stars with Gaia, but Vega is far beyond the bright limit for standard Gaia processing and they do not quantify what the uncertainties for such measurements may be. If astrometric precision for Vega were to rival the typical performance (e.g., $35\ \mu$as per measurement), we might expect Gaia to be sensitive to Jupiter masses beyond about 1.5 au \citep{Ranalli:2018}. However, we do not know the precision with which Gaia can observe a star like Vega, and it is unlikely to be close to the instrumental floor.

Radial velocities remain the most sensitive technique for companions within 1 au for the foreseeable future.

\section{Conclusions}

Using 1524 TRES spectra and two sectors of \textit{TESS} photometry, we search for planets orbiting Vega. We do not discover any transiting planets, but do detect a candidate in our radial velocities with a period of $2.43$\ days and a semi-amplitude of $6$\ \ms, implying a minimum mass of about $20$\ \mearth. Further observations and analysis will be required to confirm or refute this candidate. We use our data to derive limits on the presence of transiting planets within 0.2 au and non-transiting planets within 15 au. For orbits well aligned to the stellar spin, we are only sensitive to the most massive planets inside about 1 au, but for misaligned orbits, we are sensitive to sub-Saturn masses at small separations and the most massive planets out to about 10 au. Combining our radial velocity limits with those from previous direct imaging, we place new detection limits on brown dwarfs out to 15 au. The \textit{TESS} light curve is remarkably quiet and shows no signs of a transiting planet. With transit injection recovery tests, we can rule out most planets with radii greater than $3$\ \rearth\ and periods between 0.5 and 15 days.

We also identify rotational modulation in our data, which dominates the radial velocities but is weak in our photometry, consistent with variation driven primarily by bright plages, rather than dark spots. We model this signal with a quasi-periodic Gaussian process and with multiple Keplerians, both of which suggest that the structures on Vega's surface evolve on timescales much longer than the rotation period, implying that at least some of the surface features may be fueled by a failed fossil magnetic field.

Future high resolution spectroscopy offers a path forward for the direct or indirect detection of short-period planets orbiting Vega while simultaneously characterizing the stellar surface features and the underlying mechanisms driving them. Future photometry--–such as the extended \tess\ mission will further constrain the presence of transiting planets. At wider separations, JWST and the Nancy Grace Roman Space Telescope promise to provide new constraints on planets via direct imaging.

\acknowledgements{
We thank Tiffany Meshkat for providing the P1640 direct imaging detection limits, Vanessa Bailey for insight into observing Vega with the Nancy Grace Roman Space Telescope CGI, Alessandro Sozzetti for information regarding Gaia astrometric performance for bright stars, and Doug Gies for discussion of A star surface features. 
We gratefully acknowledge the contributions made by visiting TRES observers over the years: Robert Stefanik, Gabor F{\H u}r{\'e}sz, Bence B\'eky, Sumin Tang, Zach Berta-Thompson, Daniel Yahalomi, and Warren Brown. 
This paper includes data collected by the TESS mission. Funding for the TESS mission is provided by the NASA Explorer Program. We thank the TESS Architects (George Ricker, Roland Vanderspek, Dave Latham, Sara Seager, Josh Winn, Jon Jenkins) and the many TESS team members for their efforts to make the mission a continued success. 
This work utilized resources from the University of Colorado Boulder Research Computing Group, which is supported by the National Science Foundation (awards ACI-1532235 and ACI-1532236), the University of Colorado Boulder, and Colorado State University.}
\software{{\tt astropy} \citep{astropy:2018}, {\tt batman} \citep{Kreidberg:2015}, {\tt emcee} \citep{ForemanMackey:2012}, {\tt matplotlib} \citep{Hunter:2007}, {\tt numpy} \citep{Harris:2020}, {\tt PyMC3} \citep{Salvatier:2016}, {\tt radvel} \citep{fulton:2018a}, {\tt SOAP 2.0} \citep{Dumusque:2015}, {\tt Transit Least Squares} \citep{Hippke:2019}}
\facilities{FLWO:1.5m (TRES), \tess}

\bibliographystyle{aasjournals}

\bibliography{refs}

\end{document}